\documentclass[aps,twocolumn,floatfix,superscriptaddress,showpacs]{revtex4-1}

\usepackage{graphicx}
\usepackage{amsmath}
\usepackage{amssymb}
\usepackage{soul}
\usepackage{bbold}
\usepackage[usenames,dvipsnames, pdftex]{xcolor}

\usepackage[colorlinks, linkcolor=blue,citecolor=blue, urlcolor=blue]{hyperref}

\newcommand{\ket}[1]{| #1 \rangle}
\newcommand{\bra}[1]{\langle #1 |}

\newcommand{\figref}[1]{Fig.~\ref{#1}}

\usepackage{float}

\begin{document}

\title{Apparent slow dynamics in the ergodic phase of a driven many-body localized system without extensive conserved quantities}

\author{Tal{\'i}a L.~M. Lezama}
\affiliation{Max-Planck-Institut f\"ur Physik komplexer Systeme, 01187 Dresden, Germany}
\author{Soumya Bera}
\affiliation{Department of Physics, Indian Institute of Technology Bombay, Mumbai 400076, India}
\author{Jens H.\ Bardarson}
\affiliation{Max-Planck-Institut f\"ur Physik komplexer Systeme, 01187 Dresden, Germany}
\affiliation{Department of Physics, KTH Royal Institute of Technology, Stockholm, 106 91 Sweden}

\begin{abstract}
We numerically study the dynamics on the ergodic side of the many-body localization transition in a periodically driven Floquet model with no global conservation laws.
We describe and employ a numerical technique based on the fast Walsh-Hadamard transform that allows us to perform an exact time evolution for large systems and long times.
As in models with conserved quantities (e.g., energy and/or particle number) we observe a slowing down of the dynamics as the transition into the many-body localized phase is approached.
More specifically, our data is consistent with a subballistic spread of entanglement and a stretched-exponential decay of an autocorrelation function, with their associated exponents reflecting slow dynamics near the transition for a fixed system size.
However, with access to larger system sizes, we observe a clear flow of the exponents towards faster dynamics and can not rule out that the slow dynamics is a finite-size effect.
Furthermore, we observe examples of non-monotonic dependence of the exponents with time, with dynamics initially slowing down but accelerating again at even larger times, consistent with the slow dynamics being a crossover phenomena with a localized critical point.
\end{abstract}

\maketitle

{\it Introduction.}---One of the central nonequilibrium protocols in quantum many-body systems is periodic driving. 
Typically, if a system is ergodic, an external force will drive it out of equilibrium, heating it up to an infinite-temperature or a fully-mixed state, as conjectured by the eigenstate thermalization hypothesis~\cite{Deutsch:1991ju,Srednicki:1994dl,Rigol:2008bf}, and its variants applicable to periodically-driven, or Floquet,  systems~\cite{Alessio:2013,Lazarides:2014,Alessio:2014,Ponte:2015}. 
In contrast to ergodic systems, integrable ones fail to thermalize. One of the existing notions of quantum integrability is based on the existence of an extensive number of local (or quasilocal) conserved quantities, which prevails in the case of Floquet-integrable systems \cite{Gritsev:2017}. 
In particular, the latter avoid heating due to the constraints imposed by those conserved quantities, but nevertheless, are able to reach a well-defined synchronized state with maximal entropy, mainly described by a periodic generalized Gibbs ensemble~\cite{Lazarides:2014ap}.
There are other ways of controlling heating, encompassed by many-body localized systems (closed disordered interacting quantum systems that for sufficiently strong disorder exhibit many-body localization~\cite{Basko:2006hh,Gornyi:2005fv}). 
Such systems are integrable but stable against generic small perturbations, and can experience a many-body localization (MBL) transition taking place at finite-energy densities when tuning the disorder strength around its critical value~\cite{Nandkishore:rev,Abanin:rev,Alet:rev}. 
In particular, Floquet-MBL systems appear when adding sufficiently strong disorder to a Floquet system, with the Floquet-MBL transition tuned by the frequency or the amplitude of the drive \cite{Lazarides:2015,Ponte:2015,Ponte:2015prl,Abanin:2016,Rehn:2016}. 
In this case the problem of thermalization amounts to asking how the system synchronizes with its surroundings in order to reach its steady state. Understanding that question is not only of fundamental relevance but can also be feasibly addressed in an experimental set-up based in Floquet-engineering with ultracold atomic systems \cite{Holthaus:2016,Bordia:2017bb}.

Recently, many studies have argued that most of the ergodic phase in an MBL model is not a trivial metal but rather shows a precursor to the phase transition dominated by rare region effects, also  known as ``Griffiths effects", obstructing entanglement or transport due to localized inclusions \cite{BarLev:2015,Agarwal:2015,Potter:2015,Vosk:2015,Gopalakrishnan:2016,Znidaric:2016ia} (though other studies have questioned this conclusion~\cite{Karahalios:2009,Steinigeweg:2016,Bera:2017}). 
For systems with extensive conserved quantities, such an intermediate regime is characterized by zero DC conductivity and subdiffusive transport, as well as by a subballistic spreading of entanglement $S(t)\propto t^{1/z}$, with $z$ a disorder-dependent dynamical exponent (see \cite{Luitz:2017} for a review). 
In particular, a power-law relaxation of the density-density autocorrelation function was observed for nondriven systems with short-range interactions~\cite{BarLev:2014}, and more recently, for the spin-spin autocorrelation function in a Floquet-MBL model with conserved total magnetization~\cite{Roy:2018}.  
For systems without conservation laws, it was conjectured that in 1D the typical behavior of generic autocorrelation functions decay in time as a stretched exponential, while the average correlator still follows a power law~\cite{Gopalakrishnan:2016}; although this has never been numerically verified. 
Slower than ballistic spread of the entanglement entropy is also observed in a random circuit model~\cite{Keyserlingk:2018}.

Premised on the assumption that slow dynamics is a distinguishing feature of the ergodic phase in a class of systems exhibiting MBL when approaching the MBL transition, we show in this work that this scenario is also present in a Floquet model with no global conservation laws. 
This model is known to experience a Floquet-MBL transition which can be tuned by the disorder strength within a region of the frequency-amplitude space~\cite{Zhang:2016}. 
In particular, we study the stroboscopic dynamics of two quantities: the spin-spin autocorrelation function when the system is initially prepared in an infinite-temperature initial state, and the entanglement entropy starting from a product state. 
This setting amounts to studying how the information contained in an initial quantum state propagates in the absence of conservation laws. 
Here, the problem of thermalization is not based on the exchange of conserved quantities between the system of interest and the rest of the system but rather on the capability of the latter to undertake the necessary entanglement allowing the system to reach infinite temperature.
Using a fast Walsh-Hadamard transform, we numerically study this aspect of thermalization, for system sizes up to L = 28 and time windows extending over more than 4 decades. 

{\it Model.}---We study a kicked spin-1/2 Ising chain with open boundary conditions in the presence of both a transverse and a disordered longitudinal field, subject to a periodic driving. Its Hamiltonian can be decomposed into two terms
\begin{equation}
\begin{split}
& H_{x} = \sum_{i} \mathrm{g}\Gamma \sigma_{i}^{x},
\\& H_{z}  =  J\sum_{i=1}^{L-1}\sigma_{i}^{z}\sigma_{i+1}^{z} + \sum_{i}^{L} \left(h + \mathrm{g}\sqrt{1-\Gamma^{2}} G_{i}\right) \sigma_{i}^{z},
\end{split} 
\label{ham}
\end{equation}
where $\sigma_{i}^{x}$ and $\sigma_{i}^{z}$ are Pauli matrices on site $i$,  $G_{i}$ is a random variable with a Gaussian distribution, and the interaction constant is fixed to~$J=1$.  
The driving is induced by a time-dependent Hamiltonian that switches repeatedly between $H(t) = 2H_{z}$ and $H(t) = 2H_{x}$ every half a period $\tau/2$, so that the unitary evolution over one period is generated by the Floquet operator:
\begin{equation}
U_{F}(\tau) = e^{-iH_x\tau}e^{-iH_z\tau}
\label{fop}
\end{equation}
We  set $(\mathrm{g},h,\tau) =(0.9045,0.8090,0.8)$, as in \cite{Hyungwon:2013,Zhang:2015,Zhang:2016}, in order to make the system nonintegrable (see also~\cite{Prosen:2002,Prosen:2007} for studies of a clean version of the model). 
Both the amplitude of the transverse field and the disorder strength are controlled by tuning the parameter $\Gamma\in[0,1]$ in Eq.~\eqref{ham}, so that the total mean-square field remains independent of~$\Gamma$. 
This Floquet system is known to undergo a Floquet-MBL transition at the critical value~$\Gamma_{c} \simeq 0.3$, from the MBL phase ($\Gamma<\Gamma_{c}$) to the ergodic phase~($\Gamma>\Gamma_{c}$)~\cite{Zhang:2016} 
(see Appendix A for details on the level spacing statistics, including the clean case which was studied in detail for a class of similar set-ups in~\cite{Haldar:2018}. As well as comments on the self-dual point of the model studied in~\cite{Bertini:2018}).
%(see~\cite{SuppMat:floquetmbl} for details on the level spacing statistics, including the clean case which was studied in detail for a class of similar set-ups in~\cite{Haldar:2018}. As well as comments on the self-dual point of the model studied in~\cite{Bertini:2018}).
%

%
\begin{figure}[tb]
\flushleft
\includegraphics[width=0.484\textwidth]{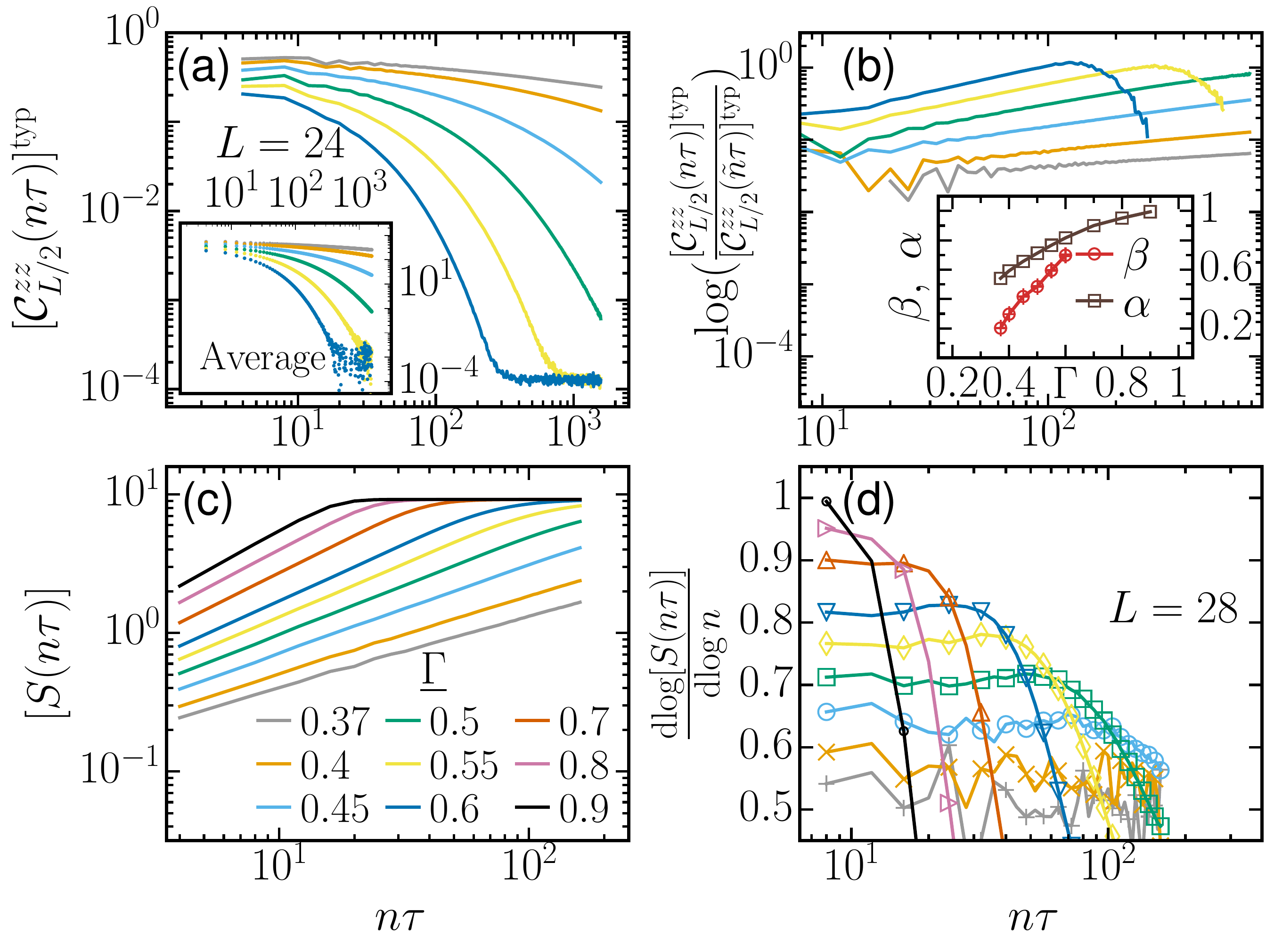}
\caption{Stroboscopic time evolution (a) of the typical disorder-averaged autocorrelation function~$[C_{L/2}^{zz}(n\tau)]^{\mathrm{typ}}$ for $L=24$, (b) the quotient between~$[C_{L/2}^{zz}(n\tau)]^{\mathrm{typ}}$ and~$[C_{L/2}^{zz}(\tilde{n}\tau)]^{\mathrm{typ}}$, measured at stroboscopic time steps $n$ of length $l_{n}$ and $\tilde{n}$ of length $l_{\tilde{n}}$, respectively (with $\tilde{n}=n/2$), for~$L=24$. (c) Stroboscopic time evolution of the disorder-averaged entanglement entropy $[S(n\tau)]$ and (d) its logarithmic derivative for~$L=28$. Data in (a), (b), (c) and (d)~correspond to the values of $\Gamma$ on the ergodic side of the transition in~(c). Inset in (a): the same as in (a) but the average. Inset in (b): Dynamical exponents~$\beta$ and $\alpha$ as a function of~$\Gamma$; the exponent~$\beta$ is extracted by fitting a linear function to the data points in~(b), whereas~$\alpha$ corresponds to the first data points in~(d).
}
\label{dqg}
\end{figure}

Here, we focus on the stroboscopic time evolution of the system, given by the Floquet operator~\eqref{fop}, for different values of disorder strength on the ergodic side of the transition~$\Gamma>\Gamma_{c}$, and as a function of system size. 
Even though~\eqref{fop} is nonintegrable, the driving is always induced by an integrable Hamiltonian, either $H_{x}$ or $H_{z}$. 
Nevertheless, the periodic switching between the two dynamics is sufficient to allow the system to absorb energy and reach an infinite-temperature state in the long-time limit 
(see Appendix B for a detailed calculation of the energy absorption).
%(see~\cite{SuppMat:floquetmbl} for a detailed calculation of the energy absorption).

%
For the chosen parameters, the model is located in a suitable region within the driving frequency-amplitude space where there is a Floquet-MBL transition as a function of disorder strength. 
Our results remain valid for a range of parameters \emph{close} to those values where we observe the same qualitative dynamical behavior of both the autocorrelation function and the entanglement entropy as a function of $\Gamma$; both heat up to infinite temperature and behave monotonically with~$\Gamma$. 

The Floquet operator~\eqref{fop} is a product of matrices that are diagonal in the respective spin basis $\sigma_x$ and $\sigma_z$.
In a diagonal basis, matrix multiplication is fast ($N=2^L$ operations as opposed to $N^2$ for a full matrix).
The basis transformation from $\sigma_z$ to $\sigma_x$ is $U = \bigotimes_{i=1}^L U_H$  with $U_H = \frac{1}{\sqrt{2}}\begin{pmatrix} 1 & 1 \\ 1 & -1\end{pmatrix}$ the $2\times 2$ Hadamard matrix.
The product structure of the transformation allows us to use the fast Hadamard transform~\cite{Fino:76}, which is a generalization of the fast Fourier transform that requires only $N\log N \sim L 2^L$ operations, to transform between the bases and thereby do an exact time evolution for large systems to large times (here we go to $L=28$ but larger systems are easily obtainable for shorter times).
\begin{figure*}[tb!]
\centering
\includegraphics[width=1\textwidth]{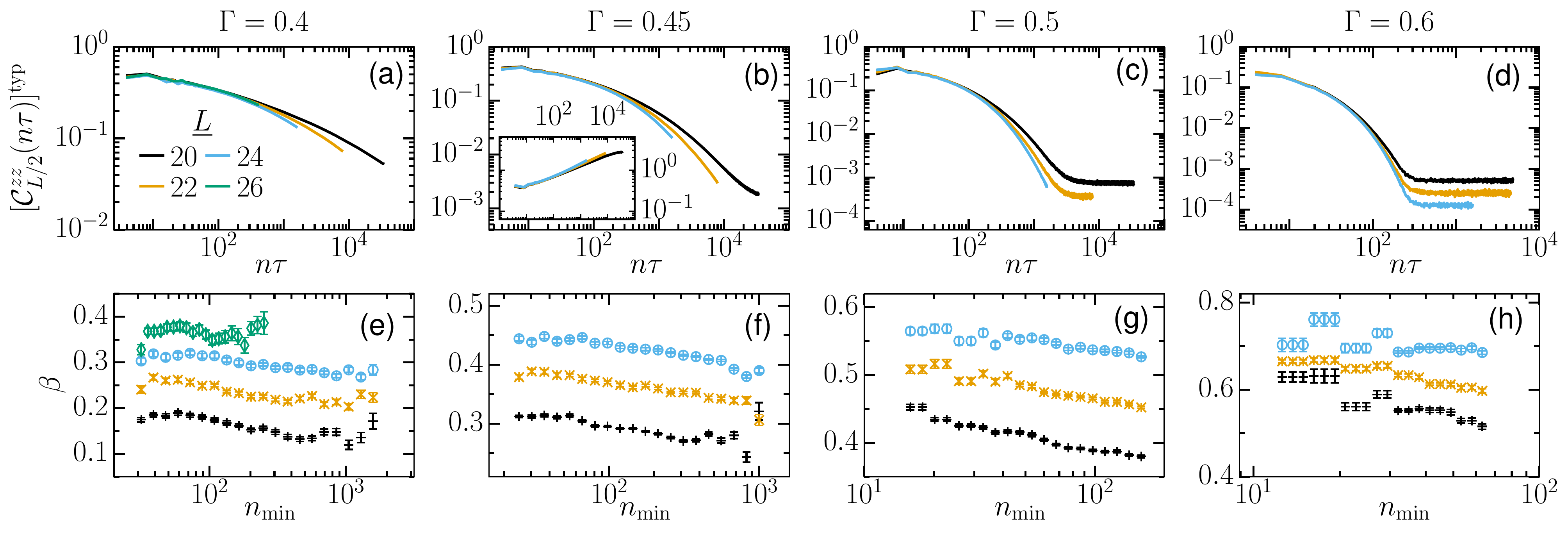}
\caption{Typical disorder-averaged stroboscopic time evolution of the autocorrelation function, $[C_{L/2}^{zz}(n\tau)]^{\mathrm{typ}}$, as a function of $L$. (a),(b)~Close to the transition; (b),(c)~deeper on the ergodic side of the transition. Inset: as in (b) but $-\log([C_{L/2}^{zz}(n\tau)]^{\mathrm{typ}})$. (e)-(h)~Dynamical exponent $\beta$ as a function of $\Gamma$~and~$L$, obtained by stretched-exponential fits to the data points in the upper panel. The fits are of fixed size and are made over $500, 400, 400$ and $200$ stroboscopic time points (from (e) to (h)), starting from~$n_{\mathrm{min}}$.
}
\label{rp}
\end{figure*}

{\it Autocorrelation function.}---For an infinite-temperature initial state the spin-spin autocorrelation function can be expressed as:
\begin{equation}
C^{zz}_{i}(t) = \frac{1}{d_{\mathcal{H}}} \mathrm{Tr}\Big(\sigma_{i}^{z}(t)\sigma_{i}^{z}(0)\Big),
\label{retpro}
\end{equation}
where $d_{\mathcal{H}}$ is the dimension of the Hilbert space. In \eqref{retpro}, we implicitly used that $\rho=\mathbb{1}$ is the associated infinite-temperature density matrix and that $\sigma_{i}^{z}(t)$ is the time-evolved operator in the Heisenberg representation; the same expression can also be thought of as a return probability. Following the notions of stochastic trace evaluation or quantum typicality \cite{Weisse:2006,Steinigeweg:2014}, the trace in \eqref{retpro} can be approximated by the expectation value with respect to $R$ initial random vectors $\{\ket{\Psi}\}$ taken from the Haar measure, up to a precision $\propto 1/\sqrt{R d_{\mathcal{H}}}$. This implies one gets converged results even for a very small number of random vectors ($R=2$). 

In what follows, using the mentioned approach, we numerically study the stroboscopic time evolution of the autocorrelation function at the middle of the lattice, $C_{L/2}^{zz}(n\tau) = \bra{\Psi}\sigma_{L/2}^{z}(n\tau)\sigma_{L/2}^{z}(0)\ket{\Psi}$, where $n$ denotes the stroboscopic time step. We averaged over $300$ to $600$ disorder realizations for each initial random vector 
(see Appendix C).
%(see~\cite{SuppMat:floquetmbl} for details regarding the trace approach).
%
In \figref{dqg}(a) we identify slow dynamics of the autocorrelation function $C_{L/2}^{zz}(n\tau)$ for values of $\Gamma$ on the ergodic side of the transition~($\Gamma>\Gamma_{c}$). 
More precisely, the autocorrelation function decays to its infinite-temperature steady state as a stretched exponential~$C_{L/2}^{zz}(n\tau) \propto \exp(-\gamma n^{\beta})$.
In \figref{dqg}(a), we show the typical and the average (inset) value of~$C_{L/2}^{zz}(n\tau)$, observing a faster decay of both with increasing~$\Gamma$ (decreasing disorder strength) for a fixed $L$. The average is denoted as~$[\cdot]$ and the typical $\exp\left([\log(\cdot)]\right)$ as $[\cdot]^\mathrm{typ}$.
While we observe a stretched-exponential behavior of the typical value of the autocorrelation function, we found no difference with the average, and consequently, no subleading power-law behavior of the latter, as was conjectured in~\cite{Gopalakrishnan:2016} for generic autocorrelation functions in 1D systems without extensive conserved quantities. 

In~\figref{dqg}(b), we further show how the decay of the autocorrelation function reflects on the value of the dynamical exponent~$\beta$ for a fixed system size~$L$ and the same in~\figref{rp} as a function of~$L$. 
We extracted the exponent in two ways: 
The first one consists in taking the logarithm of the quotient between the autocorrelation function~$C^{zz}_{L/2}(n\tau)$ measured at stroboscopic times~$n\tau$ of length~$l_{n}$, and the autocorrelation function~$C^{zz}_{L/2}(\tilde{n}\tau)$ measured at stroboscopic times~$\tilde{n}\tau$ of length~$l_{\tilde{n}} = 2l_{n}$, with~$\tilde{n}=n/2$. This is shown in~\figref{dqg}(b) in a log-log scale.
As both the numerator and denominator in  $\log\Big(\frac{C^{zz}_{L/2}(n\tau)}{C^{zz}_{L/2}(\tilde{n}\tau)}\Big)$ decay as a stretched exponential, by taking again the logarithm we can therefore extract the dynamical exponent  $0.2 \lesssim \beta \lesssim 0.7$ via a linear fit $\propto \delta+ \beta (n\tau)$, plotted in the inset. 
The exponent grows with increasing~$\Gamma$, and is in good agreement with the one alternatively extracted via stretched-exponential fits plotted in the lower panel of~\figref{rp}. 
The fits are made over several time windows of fixed size~$\Delta n$ for a given~$\Gamma$; starting at different initial stroboscopic times~$n_\mathrm{min}$. 
Both the order of~$\Delta n$ and~$n_\mathrm{min}$ are naturally constrained by the finite-size effects which increase with~$\Gamma$. 
The exponent corresponding to the stretched-exponential fits,~$0.3 \lesssim \beta \lesssim 0.7$, grows with increasing~$\Gamma$ and fluctuates only slightly about those approximate values with respect to~$n_\mathrm{min}$, for a given~$L$ (see data for $L=24$ to compare with the corresponding data shown in the inset of \figref{dqg}(b)).

\begin{figure*}[tb!]
\centering
\includegraphics[width=1\textwidth]{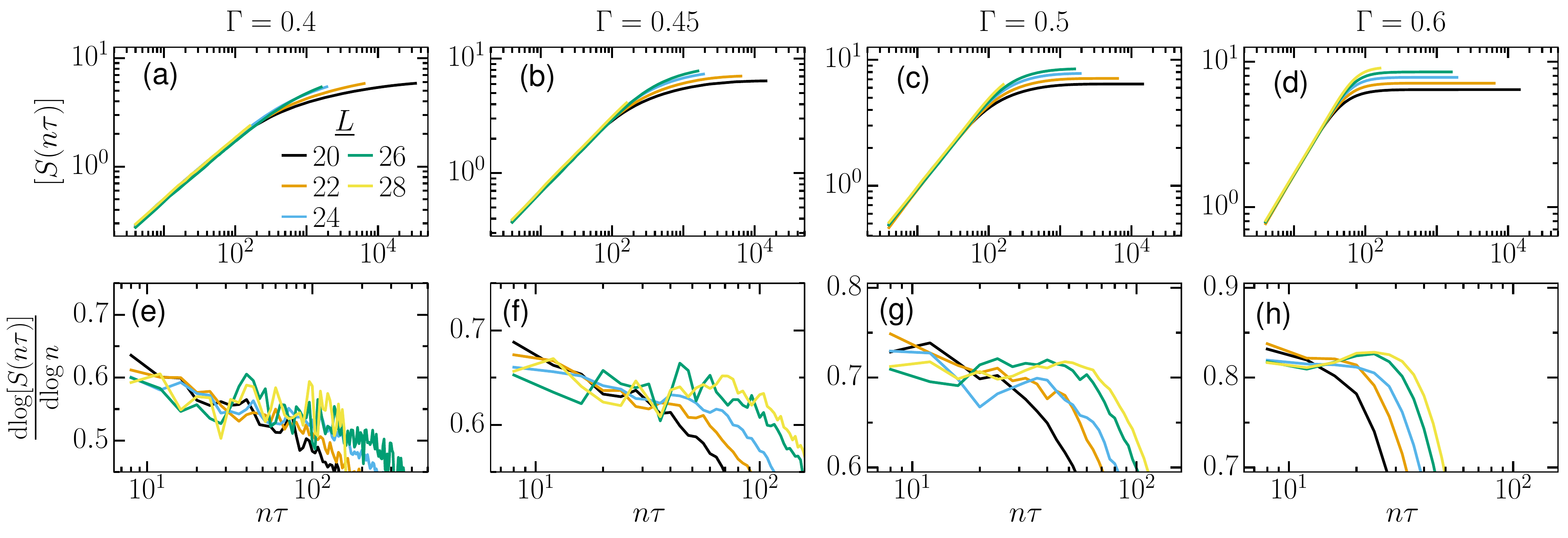}
\caption{Disorder-averaged stroboscopic time-evolution of the entanglement entropy, $[S(n\tau)]$, as a function of $L$. (a),(b) Close to the transition; (c),(d) deeper on the ergodic side of the transition. (e)-(h) The logarithmic derivative of the data points in the upper panel. The power-law regime is signaled by the plateaus, whose range increases with $L$ for a given $\Gamma$. The dynamical exponent $\alpha$ can be approximately obtained within that range as a function of $\Gamma$, observing that the range of the power-law regime decreases with increasing $\Gamma$ (decreasing disorder strength) resembled on the saturation rate of $S(n\tau)$.
}
\label{ent}
\end{figure*}

For a fixed $\Gamma$, the value of the dynamical exponent~$\beta$ increases with~$L$, then flowing towards faster dynamics for larger system sizes (see lower panel of~\figref{rp}).
So, although we observe a slow dynamics as a function of disorder within the ergodic regime, we also observe that the slow dynamics fasten with system size. 
Both the presence or absence of slow dynamics could be related to the small system sizes reached in our simulations. Either the system sizes explored here are too small to attest that the exponent saturates at a value corresponding to such slow dynamics for a larger system size, or the slow dynamics is a direct consequence of the finite-size effects. Recently, the former scenario was observed in a nondriven system, where a converged exponent was obtained but only until system sizes $L \sim 50-100$ were attained~\cite{Doggen:2018}; a large crossover length scale at weak disorder was discussed in in Ref.~\onlinecite{Znidaric:2016ia}, also in the nondriven setting.

{\it Entanglement entropy.}---As we stated before, the system described in \eqref{ham} does not possess any conservation laws; while there is no diffusive transport in the conventional hydrodynamical sense \cite{Zotos:1997}, there is still transport of quantum information. It is therefore interesting to explore how the absence of conserved quantities influences the spread of the entanglement entropy,
\begin{equation}
S(t)=-\mathrm{Tr} [\rho_{A}(t)\ln(\rho_{A}(t))],
\end{equation}
where $\rho_{A}(t) = \mathrm{Tr}_{B}[\rho(t)]$ is the time-evolved reduced density matrix for a bipartite system $(A|B)$ and $\rho=\ket{\Psi}\bra{\Psi}$ the density matrix for a given initial product state $\ket{\Psi}$. 
Typically, for 1D MBL systems on the ergodic side of the transition, the entanglement growth after a quench from a product state is subballistic $S(t)\propto t^{\alpha}$, where~$\alpha\leq 1$ is a disorder-dependent dynamical exponent with the upper limit given by the Lieb-Robinson bound \cite{Luitz:2016,Lieb1972}. However, such entanglement spread has been mostly explored in MBL systems with extensive conserved quantities (with the exception \cite{Keyserlingk:2018}).

In \figref{dqg}(c), we plot the stroboscopic time evolution of the entanglement entropy $S(n\tau)$ after a quench from a N{\'e}el state, for several values of $\Gamma$ on the ergodic side of the transition. 
As in the case of systems with conserved quantities, we observe a subballistic entanglement growth in terms of the stroboscopic time steps~$n$, $S(n\tau)\propto (n)^{\alpha}$. 
The rate of entanglement saturation towards the page value~$L/2\log(2)-0.5$ is much faster compared with the one of the undriven model (data not shown). This is in agreement with~\cite{Hyungwon:2013}, where the same observation is made for the clean version of the model.

In order to extract the exponent $\alpha$ for a fixed $L$ as a function of $\Gamma$, we calculate the logarithmic derivative of the entanglement entropy (see \figref{dqg}(d)) and then take the first points where we see a plateau in time, plotted in the inset in \figref{dqg}(b). 
We find an exponent $0.5 \lesssim \alpha \lesssim 1$ that decreases with decreasing~$\Gamma$; even if the upper limit could be extracted from hardly a plateau, it is consistent with a flow towards a ballistic spreading of entanglement when approaching the clean case ($\Gamma=1$).  

While the exponent~$\beta$ seems to flow to zero when approaching the critical point~$\Gamma_{c}\approx 0.3$, we do not see that the exponent~$\alpha$ goes to zero, presumably because when approaching the critical value~$\Gamma\approx\Gamma_{c}$, once the power becomes~$\alpha\approx 0.5$, it is hard to distinguish a logarithm (the expected entanglement growth behavior in the MBL phase \cite{Znidaric:2008,Bardarson:2012gc}) from a power law, so we can not reliably extract the exponent using the analysis exposed above for smaller values of~$\Gamma$. 
We also explore the spreading of entanglement and its logarithmic derivative as a function of system size~$L$, both plotted in the upper and lower panel of~\figref{ent}, respectively. 
As above, the power-law regime is signaled by the plateaus in time, whose extent grows with $L$ and reduces with increasing $\Gamma$. 
Furthermore, when looking carefully at the lower panel of \figref{ent}, we observe a non-monotonic dependence of the exponents; the dynamics starts slowing down and then accelerates with increasing time, for a fixed $\Gamma$, seen as an upturn of $\alpha$ at longer time and at larger $L$.
This suggests that the slow dynamics might be a transient phenomena (as also observed in certain translationally invariant lattice models~\cite{Michailidis:2018}).

{\it Discussion.}---In summary, using the fast Hadamard transform, we have been able to study dynamics on the ergodic side of the transition in a Floquet model of many-body localization in large systems sizes (up to $L = 28$) and large times ($n\tau > 10^4$).
While for a given system size we observe clear slow dynamics, reflected in stretched exponential decay of the autocorrelation function and subballistic spreading of the entanglement entropy, this dynamics consistently speeds up with increasing system size.
Large system sizes even allow us to observe examples of dynamics that being initially slow, at later time, speed up before reaching saturation, reminiscent of what is observed in large scale simulations of random regular graphs~\cite{Tikhonov:2016}. 
Such behavior would, for example, be consistent with a localized critical point and the initial dynamics for small systems in the ergodic phase still being under the influence of the many-body critical point, but later flowing away from it into fully ergodic dynamics.
Alternatively, the speeding up of the dynamics could be just a finite-size effect, and at sufficiently large systems (which we can not reach), the flow of exponents would saturate at non-ergodic values. 
While we can not decide between the two options based on our numerical data, our conclusions show that the former scenario is well consistent with our results, which would mean the absence of a Griffiths phase in the model we study. 
\\

\acknowledgements
%{\it Acknowledgements.}---
We thank Henning Schomerus for helping us figuring out the Hadamard-Walsh transform and for useful comments on the manuscript; we further thank Yevgeny Bar Lev, Achilleas Lazarides, and David Luitz for many valuable discussions throughout the elaboration of this work. We extend thanks to Hadi Yarloo for stimulating discussions. This work was supported by the ERC Starting Grant No.~679722 and the Knut and Alice Wallenberg Foundation 2013-0093. SB acknowledges support from DST, India, through Ramanujan Fellowship Grant No. SB/S2/RJN-128/2016.

\bibliography{references}

\appendix

\section{Level spacing statistics}

In order to make the paper self-contained, in this appendix we provide details on the level statistics in the model studied using exact diagonalization over $100-10^{4}$ disorder realizations. The Floquet operator defined in Eq.~\eqref{fop} of the main text, has eigenvalues of the form $e^{-i\theta_{n}}$, the phases~$\theta_{n}$ being directly related to the quasienergies ($\varepsilon_{n}=\theta_{n}\tau$), and defined in the interval~$(-\pi,\pi ]$. Given the ordered phases $\theta_{n+1}\geq \theta_{n}\geq \dots \geq \theta_{1}$, the level spacing ratio between two consecutive phase spacings $\delta_{n}$ is defined as~\cite{Alessio:2014}
\begin{equation}
r = \frac{\mathrm{min}(\delta_{n}, \delta_{n+1})}{\mathrm{max}(\delta_{n},\delta_{n+1})}; \quad \mathrm{with} \quad  \delta_{n} = \theta_{n+1}-\theta_{n}.
\label{eq:lsr}
\tag{S1}
\end{equation}

\begin{figure}[tb]
\centering
\includegraphics[width=0.45\textwidth]{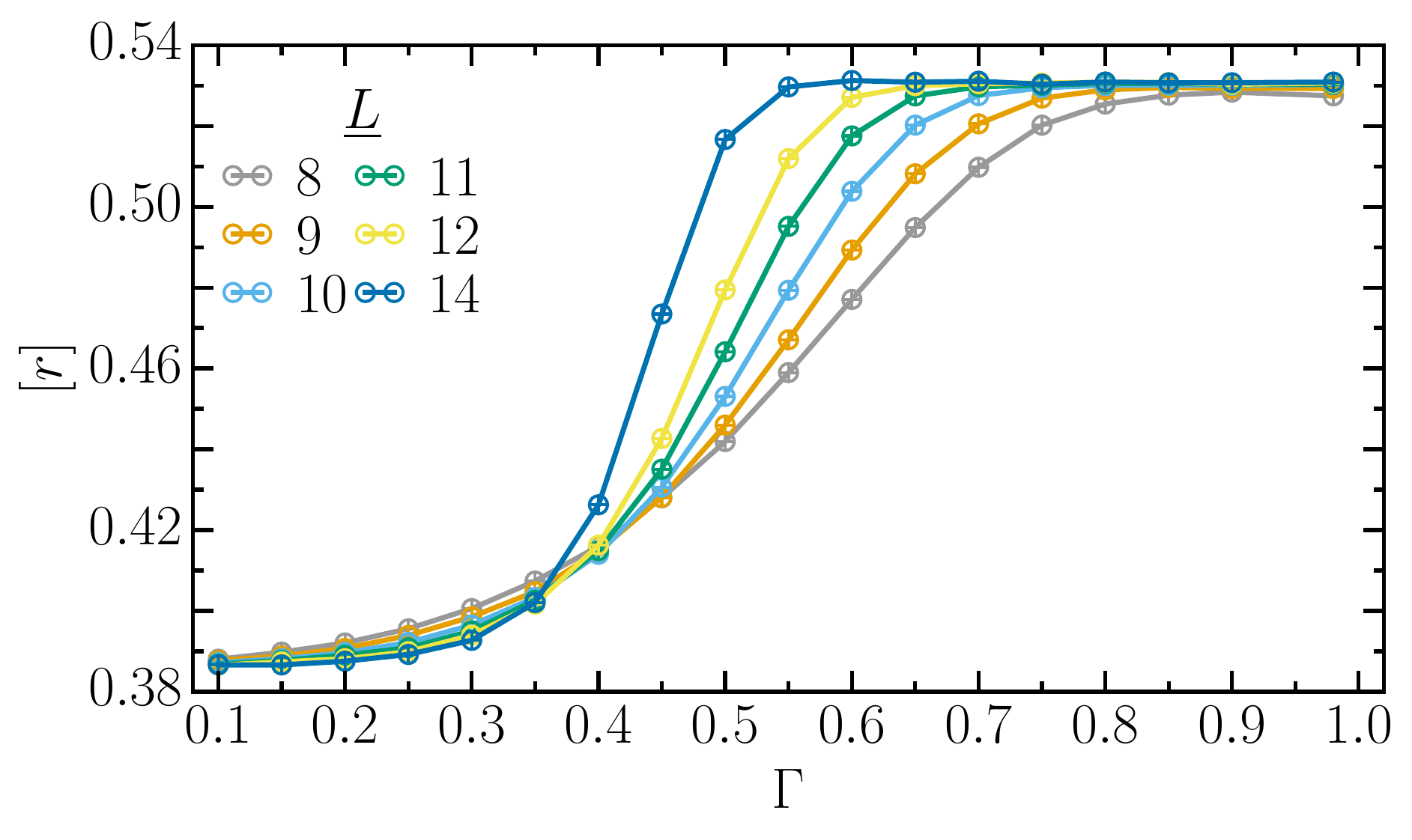}
\caption{Average value of the level spacing ratio, $[r]$, as a function of $\Gamma$, for several system sizes $L$. 
}
\label{lsr}
\end{figure}

In~\figref{lsr} we plot the average value of the level spacing ratio,~$[r]$, as a function of $\Gamma$. We note that~$[r]$ is enclosed by the limit values corresponding to the Poisson~(POI) and the circular ortogonal ensemble~(COE) distributions,~$[r]_{\mathrm{POI}} \approx 0.386$ and $[r]_{\mathrm{COE}} \approx 0.526$, respectively. The critical value of disorder is signalled by the crossings and is located below~$\Gamma\approx0.35$ for the system sizes reached in our simulations. As the crossing shifts towards $[r]_{\mathrm{POI}}$ with increasing $L$, the data shown in~\figref{lsr} is well consistent with the critical disorder $\Gamma_{c}\approx 0.3$ reported in~\cite{Zhang:2016}, where the same quantity is featured for exactly the same model. 

\begin{figure*}[tb!]
\centering
\includegraphics[width=0.95\textwidth]{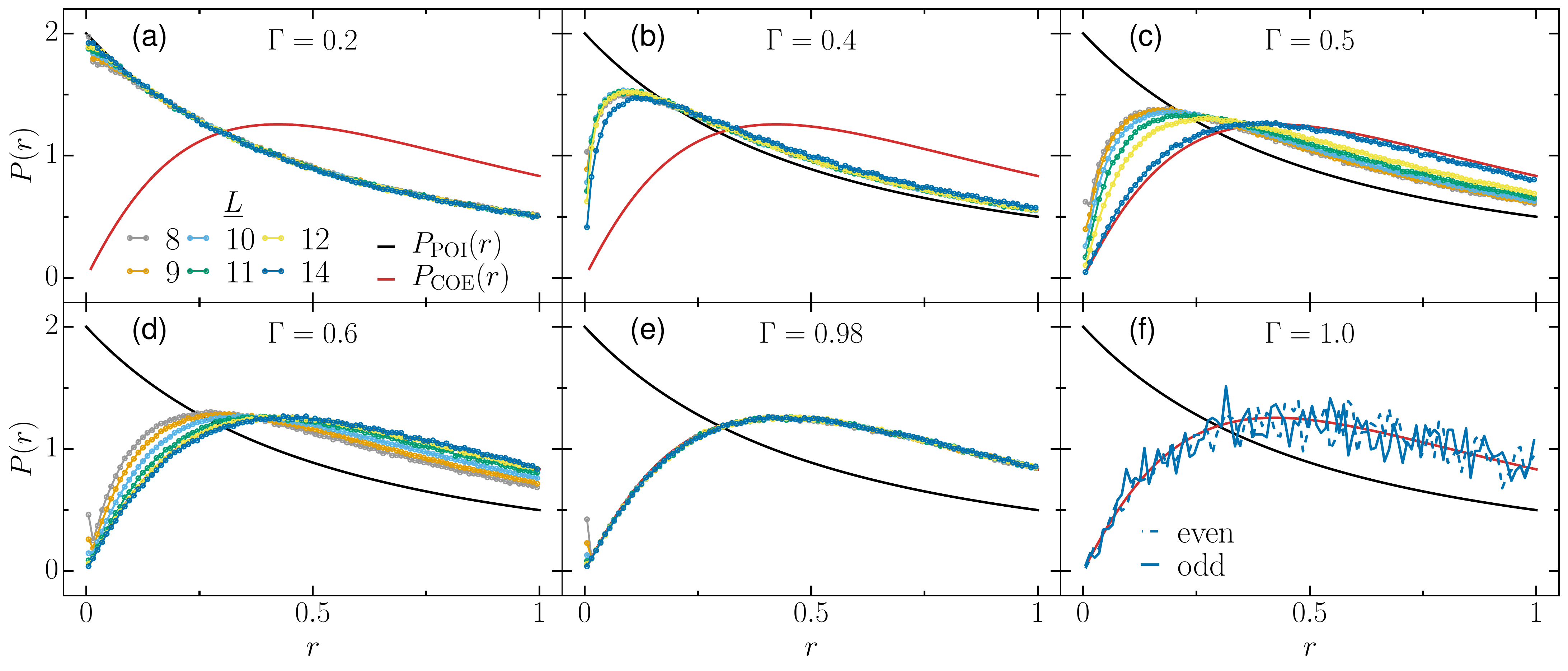}
\caption{Probability distribution $P(r)$. (a) Deep in the localized phase. (b) Close to the transition; (c)-(e) deeper on the ergodic side of the transition. (e) In the disorder-free case for the even (dotted lines) and odd (solid lines) parity sectors, for L=14. 
}
\label{pdis}
\end{figure*}

We further plot the full probability distribution of $r$, $P(r)$, for different values of disorder (see~\figref{pdis}). For values of $\Gamma$ deep in the MBL phase ($\Gamma$ close to zero), we expect $P(r)$ to obey the Poisson distribution~\cite{Atas:2013}
\begin{equation}
P_{\mathrm{POI}}(r) = \frac{2}{(r+1)^{2}}.
\label{eq:poid}
\tag{S2}
\end{equation}

While for values of $\Gamma$ deep in the ergodic phase ($\Gamma$ close to one), we expect $P(r)$ to obey the COE distribution~\cite{Alessio:2014}
\begin{equation}
\begin{split}
P_{\mathrm{COE}}(r) =& \frac{2}{3}\left(\frac{\sin\left(\frac{2\pi r}{r+1}\right)}{2\pi r^{2}}+\frac{\sin\left(\frac{2\pi}{r+1}\right)}{2\pi}+\frac{1}{(r+1)^{2}}\right. \nonumber \\&\left. -\frac{\cos\left(\frac{2\pi r}{r+1}\right)}{r(r+1)}-\frac{\cos\left(\frac{2\pi}{r+1}\right)}{r+1}\right).
\end{split}
\label{eq:pcoe}
\tag{S3}
\end{equation}
In~\figref{pdis} we compare our numerical results with the two previous closed expressions. In particular, the limit cases of very strong disorder ($\Gamma=0.2$) and very weak disorder ($\Gamma=0.98$) are plotted in~\figref{pdis}(a),(e) which show that~$P(r)$ is indeed well described by Eq.~\eqref{eq:poid}~and Eq.~\eqref{eq:pcoe}, respectively. It is interesting to observe though, how the distributions develop for intermediate disorder strengths. While close to the critical point, the distributions seem to be independent of~$L$ (see \figref{pdis}(b)), when moving away from the critical point, the distributions seem to display strong finite size-effects. This latter observation can be seen in detail in~\figref{pdis}(c),(d), where $P(r)$ seems to match $P_{\mathrm{COE}}(r)$ quite well already for~$L=14$, but not for $L<14$. 

Furthermore, even if the model considered is ergodic in a vicinity of the disorder-free point $\Gamma=1.0$ (see~\figref{pdis}(e)), it is relevant to confirm that the model remains ergodic in the disorder-free case, where ergodicity has been observed in~\cite{Haldar:2018} for a class of clean similar set-ups. In this case, the model is invariant under reflection, leading to parity conservation, i.e.,~$[U_{F}(\tau),\hat{\Pi}] =0$, where $\hat{\Pi}$ is the parity operator~\cite{Kira:2013}
\begin{equation}
 \hat{\Pi} =\prod_{i=1}^{\tilde{L}}\frac{1}{2} \left(\sigma_{i}^{x}\sigma_{L-i+1}^{x}+\sigma_{i}^{y}\sigma_{L-i+1}^{y}+\sigma_{i}^{z}\sigma_{L-i+1}^{z} + \mathbb{1}\right);
\label{pop}
\end{equation} 
with $\tilde{L}=\frac{L}{2}$ for $L$ even and $\tilde{L}=\frac{(L-1)}{2}$ for $L$ odd. 

If we were to inspect the level statistics in the disorder-free case, we necessarily have to consider the even and odd parity sectors ($\hat{\Pi} =\pm 1$) separately. In~\figref{pdis}(f), we show that both parity sectors are ergodic. Our results confirm that a single large enough system, such as~$L=14$, is already sufficient to reasonably reproduce the COE ensemble. Alternatively, we could have broken the symmetry by adding a small random coupling in the border of the chain (data not shown).

{\bf The self-dual point.}---
A recent analytical connection to random matrix theory in terms of the spectral form factor was derived in~\cite{Kos:2018}. In particular, this analytical tool was used to study ergodicity at the self-dual point of basically the same model considered here, leading to a theorem that states the non-existence of MBL regardless of the disorder strength in the system~\cite{Bertini:2018}.

The self-dual point in our model corresponds to setting $J\tau = \frac{\pi}{4}$ and $g\Gamma\tau = \frac{\pi}{4}$ in Eq.~\eqref{ham} of the main text.
Using the rudimentary measures of ergodicity in terms of probability distributions of the level spacing ratio, we reproduce the aforementioned theorem at the self-dual point. Our numerics show that the self-dual point is indeed an exact ergodic point, reflected by the COE-like distributions obtained for all the values of disorder strength explored, including those that would correspond to POI-like distributions at parameters away from the self-dual point.

\begin{figure}[tb]
\centering
\includegraphics[width=0.45\textwidth]{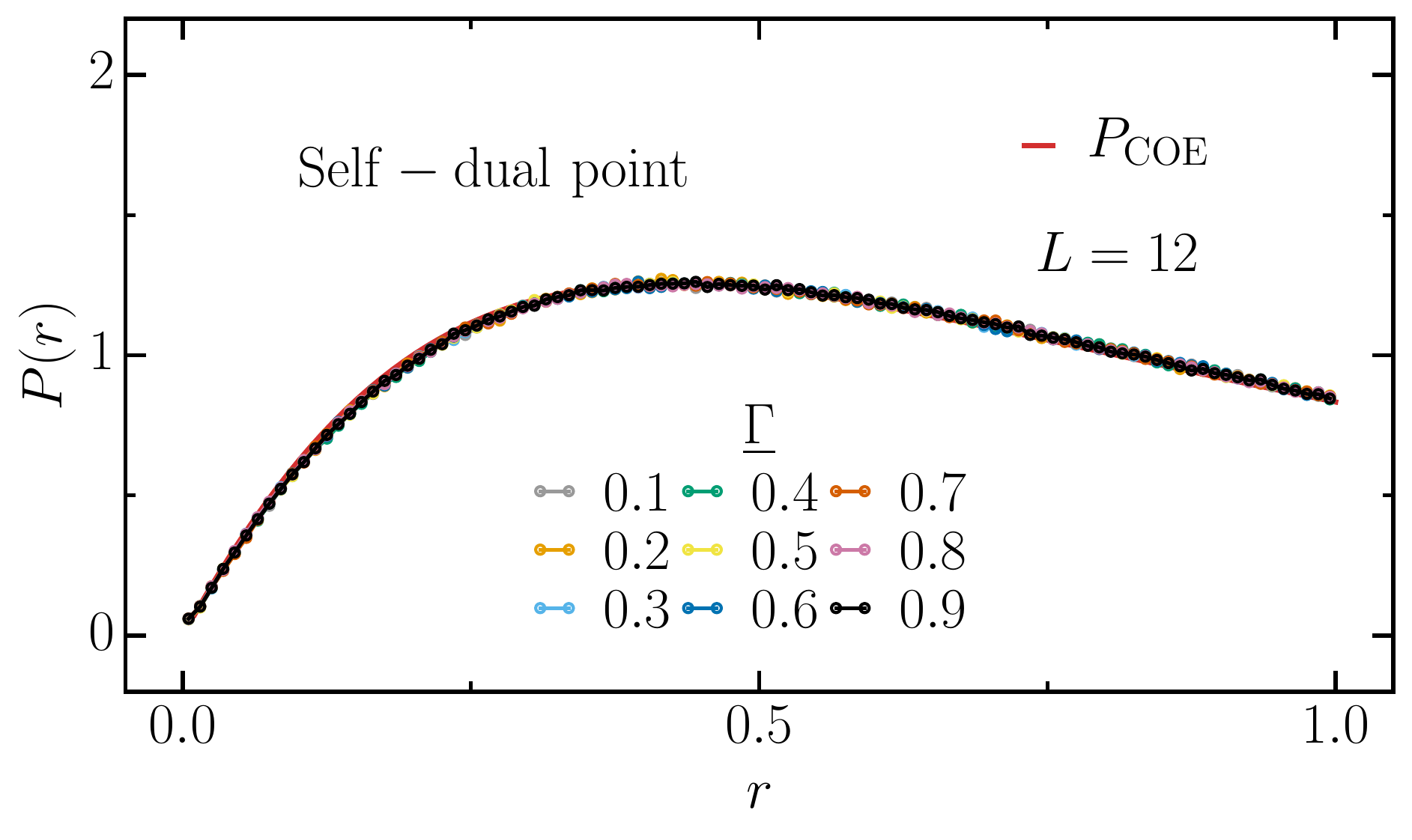}
\caption{Probability distribution $P(r)$ at the self-dual point $J\tau = g\Gamma\tau = \pi/4$ for $L=12$ and several values of $\Gamma$, including those that would correspond to the MBL regime at other points of the parameter space.
}
\label{sdp}
\end{figure}

\section{Energy absorption}

As mentioned in the main text, it is already well established that the fate of a  static system in its MBL phase when subjected to periodic driving, will depend on both the period and the amplitude of the drive.  If the amplitude is small enough, then, for fast enough driving the system will remain localized, whereas for slow driving the system will delocalize to its ergodic phase. Both scenarios can be reflected in how the energy is absorbed in real time. For fast enough driving, the energy of the system will remain localized in the infinite-time limit; for slow driving, the system will heat up to infinite temperature~\cite{Alessio:2013,Alessio:2014,Rehn:2016}. 

The analysis presented in this work slightly differs from the aforesaid scenarios, as the model considered here has a fixed driving period which allows for a Floquet-MBL transition tuned by the disorder strength. Nevertheless, here, we expect to similarly observe a distinguishing heating process on each side of the transition. Following the same reasoning as in~\cite{Alessio:2014,Rehn:2016}, we periodically drive the system initially prepared in the ground state of the time-averaged Hamiltonian
\begin{equation}
H_{\mathrm{avg}}\equiv\frac{1}{\tau}\int_{0}^{\tau}dtH(t) = H_{x}+H_{z},
\label{havg}
\end{equation} 
to then study how the energy is absorbed in real time. The stroboscopic time evolution of the energy density with respect to $H_{\mathrm{avg}}$ is defined as
\begin{equation}
\varepsilon(n\tau) =  \bra{\Psi} H_{\mathrm{avg}}(n\tau)\ket{\Psi},
\label{eq:eabs}
\end{equation}
with $n$ the stroboscopic time step and $\tau$ the period used in the main text. If the system heats up to infinite-temperature in the long-time limit, then the energy absorbed~\eqref{eq:eabs} at long times is
\begin{equation}
\varepsilon(n\tau\rightarrow\infty) = \mathrm{Tr}\left(H_{\mathrm{avg}}\right) = 0,
\end{equation}
where the last equality follows from Eq.~\eqref{havg} and Eq.~\eqref{ham} of the main text. Therefore, if the system heats up to infinite-temperature, then the energy absorption saturates to zero. 

\begin{figure}[htp]
\centering
\includegraphics[width=0.45\textwidth]{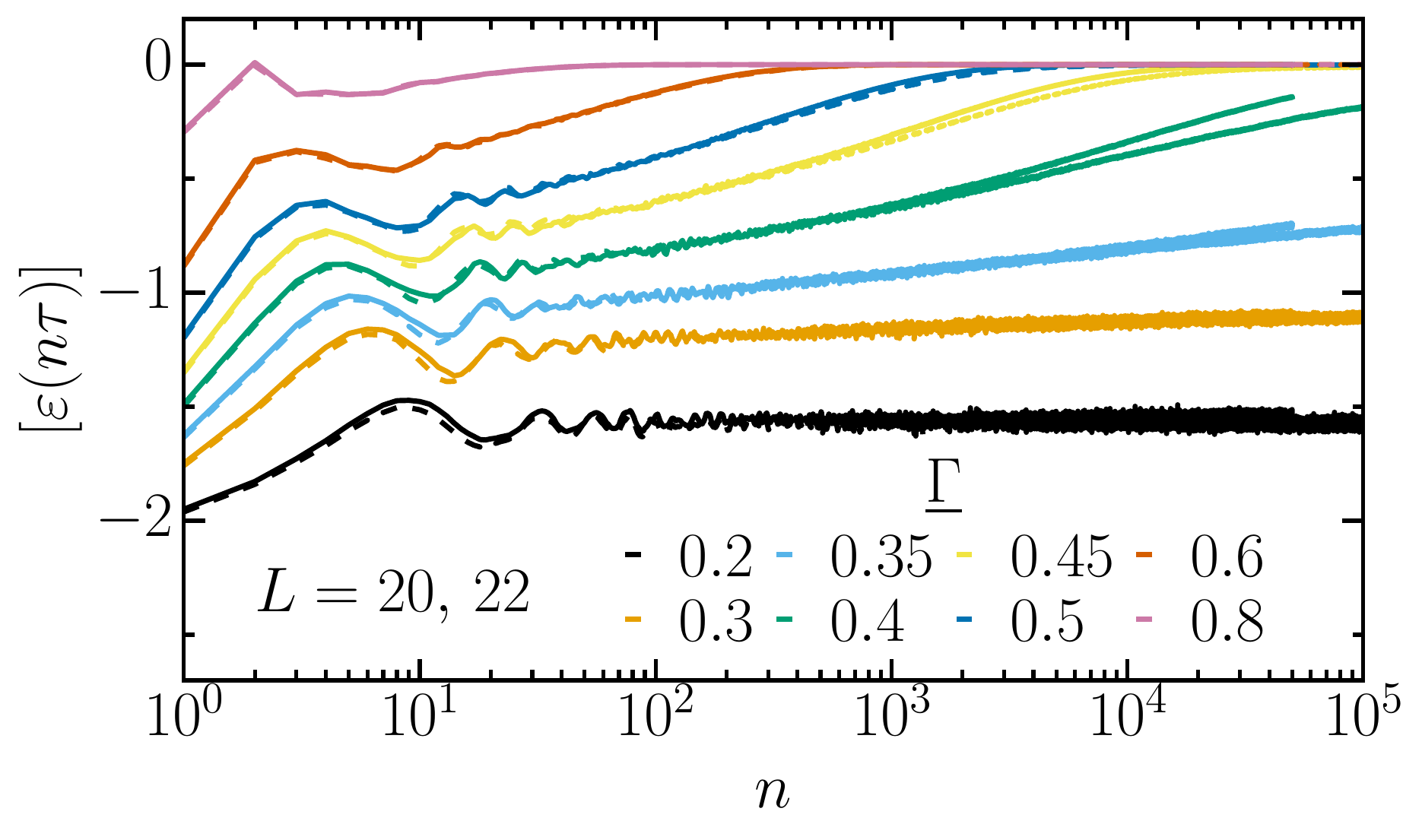}
\caption{Disorder-averaged energy absorption in the MBL phase ($\Gamma=0.2$), across and near the transition ($\Gamma=0.3,0.35$), and deeper on the ergodic side of the transition ($\Gamma=0.4,0.45,0.5,0.6,0.8$). For two system sizes; $L=20$ (solid lines), L=22 (dashed lines). 
}
\label{eabs}
\end{figure}

In~\figref{eabs} we show the energy density $\varepsilon(n\tau)$ for several values of $\Gamma$ and corroborate the existence of qualitatively different regimes characterized by different heating process.
In the localized regime, the energy remains localized in the inifinite-time limit, away from the infinite-temperature saturation point, while in the ergodic regime the system continues absorbing energy until heating up to infinite-temperature, albeit rather slowly when approaching the critical point. 
Both regimes are consistent with the phases of $U_{F}(\tau)$ being either POI or COE distributed--or falling somewhere between the two--as shown in~\figref{pdis}.

Although we observe a slow heating process between the localized and ergodic regimes, whether this is consistent with the logarithmically slow process observed for a model with conserved total magnetization in~\cite{Rehn:2016}, is not entirely clear. It might well be that the slow behavior is a finite-size effect. 

\vspace{1.5cm}

\vspace{1.5cm}

\section{Trace approximation}
In this section we provide numerical evidence showing that the trace estimate used in our simulations converges to the exact trace evaluation in Eq.~\eqref{retpro} of the main text, for a modest number of both disorder realizations and random vectors.

In \figref{tracecomp}, we plot the stroboscopic time evolution of the autocorrelation function obtained using both the exact trace evaluation and the trace approach based on the stochastic trace evaluation method. The latter consists in estimating the trace by an average over $R$ random vectors~$\{\ket{r}\}$. The statistical expectation value~$\langle \hat{A}\rangle =\mathrm{Tr}(\rho\hat{A})$ of an Hermitian operator $\hat{A}$ for a given ensemble with density matrix $\rho$, can be therefore expressed by the average over random states 
$$\langle \hat{A} \rangle \approx \sum_{r=0}^{R-1} \bra{r} \rho \hat{A} \ket{r}.
$$
As already mentioned in the main text, the relative error of the trace estimate is of order $O(1/\sqrt{R d_{\mathcal{H}}})$. In our case, $d_{\mathcal{H}} \sim 2^{L}$. While the results for a single realization start converging when increasing the number of random vectors, the disorder average significantly improves the convergence and a small number of random vectors is enough to obtain converged results for small system sizes~($L=8,10$). See upper and lower panel of \figref{tracecomp}.

\begin{onecolumngrid}
\begin{center}
\begin{figure}[htp]
\centering
\includegraphics[width=1\textwidth]{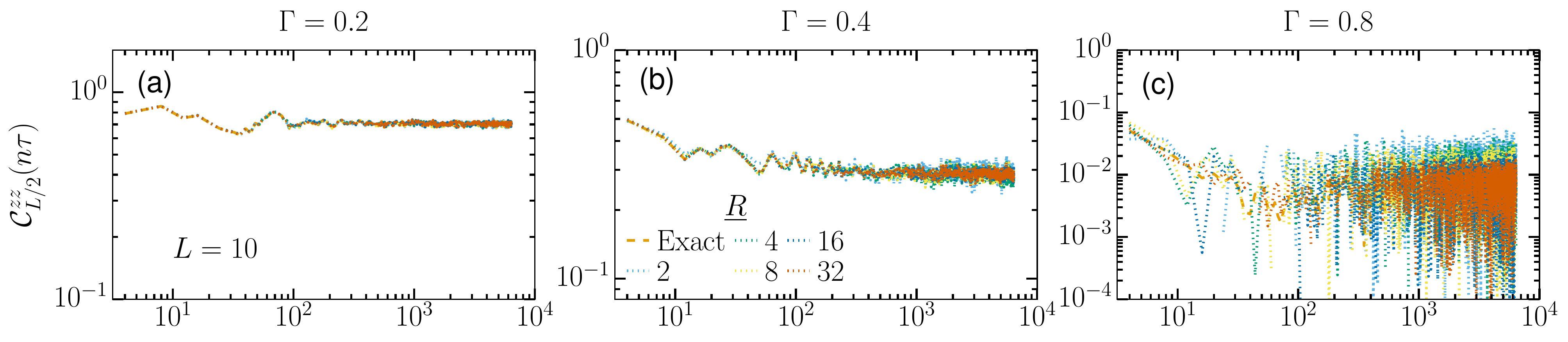}
\includegraphics[width=1\textwidth]{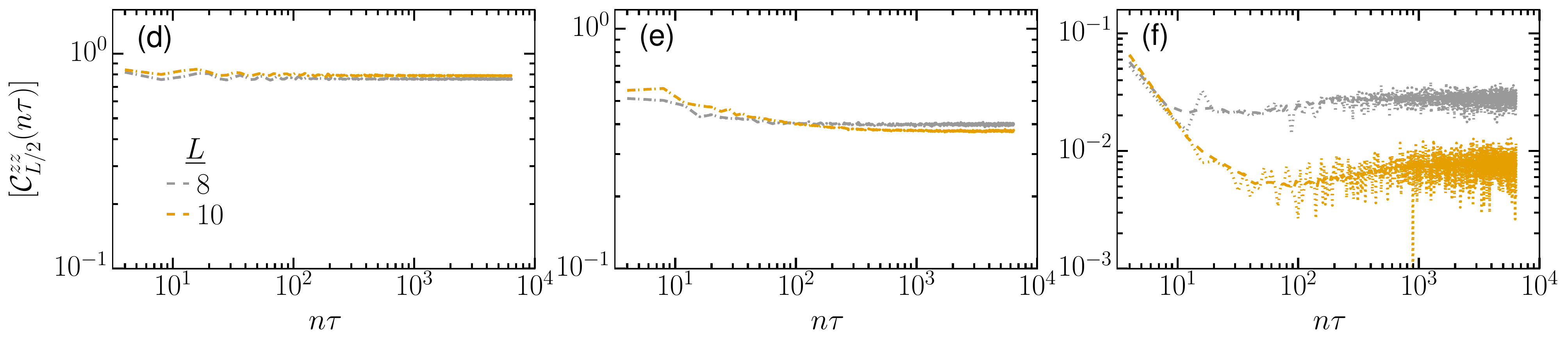}
\caption{
Stroboscopic time evolution of the return probability $C_{L/2}^{zz}(n\tau)$,  obtained using both the exact trace evaluation (dashed lines) and the stochastic trace evaluation (dotted lines). Upper panel: for a single disorder realization using different number of random vectors $R=2,4,8,16,32$~($L=10$). (a) In the localized phase ($\Gamma=0.2$), (b) on the ergodic side of the transition ($\Gamma=0.4$), and (c) deeper on the ergodic side of the transition ($\Gamma=0.8$). Lower panel: as in the upper panel but disorder averaging $[\cdot]$; we used the same 100 disorder configurations for both the exact and the estimate trace evaluation ($L=8,10$ in (d)-(f)).
}
\label{tracecomp}
\end{figure}
\end{center}
\end{onecolumngrid}

\end{document}